\documentclass[fleqn,twoside]{article}
\usepackage{espcrc2}

\usepackage{graphicx}
\usepackage[figuresright]{rotating}

\newcommand{\AmS}{{\protect\the\textfont2
  A\kern-.1667em\lower.5ex\hbox{M}\kern-.125emS}}

\title{Neutrino Telescopes}

\author{J. J. Hern\'andez--Rey\\
IFIC - Instituto de F\'{\i}sica Corpuscular,\\ 
Universitat de Val\`encia--C.S.I.C., Apdo. 22085, E-46071 Valencia, Spain}

\begin{document}

\begin{abstract}
The observation of high energy cosmic neutrinos can shed light on the
astrophysical sites and mechanisms involved in the acceleration of
protons and nuclei to the high energies observed at Earth by cosmic
ray detectors. More generally, high energy neutrinos can be a key
instrument in the multimessenger study of the high energy sky. Several
neutrino telescopes of different sizes and capabilities are presently
taking data and projects to further increase their size and
sensitivity are underway.

In this contribution we review the present status of neutrino
telescopes based on the cherenkov ligh detection technique, their
recent results and the plans to increase their sensitivity.
\vspace{1pc}
\end{abstract}

\maketitle

\section{Introduction}

The detection of high energy cosmic neutrinos is a long and much
sought-after scientific goal. In the low energy domain (few MeV to
several GeV) the observation of extraterrestrial and atmospheric
neutrinos gave rise to the discovery of neutrino oscillations and to
the -arguably- first experimental test of our models of supernova
explosion. In the high energy regime (several GeV to EeV), the
advantages of neutrinos as cosmic messengers are well known, their
importance to provide information on the particle acceleration
mechanisms in astrophysical objects is also acknowledged and
experimental methods to detect them exist and have been
technologically proven.  Therefore, at present there is a struggle to
reach the required sensitivities to detect them, the hope being that
nature is generous enough to provide fluxes at a level observable with
our present or soon-to-come neutrino telescopes.
 
Let us briefly summarize the advantages of neutrinos as cosmic
messengers. They are neutral particles, they are not deflected by
magnetic fields and therefore they point back to their sources. They
are weakly interacting and thus they can escape from very dense
astrophysical objects and can travel long distances without being
absorbed by matter or background radiation.  Moreover, in cosmic sites
where hadrons are accelerated it is likely that neutrinos are
generated in the decay of charged pions produced in the interaction of
those hadrons with the surrounding matter or radiation, being
therefore a smoking gun of hadronic acceleration mechanisms.

The observation of neutrinos in a cherenkov neutrino telescope is
based on the detection of the muons produced by the neutrino charged
current interactions with the matter surrounding the telescope by
means of the cherenkov light induced by the muons when crossing the
detector medium, natural ice or water. The telescope consists in a
three dimensional array of light sensors, usually photomultipliers,
that record the position and time of the emitted cherenkov photons and
therefore enable the reconstruction of the muon track.  To avoid the
huge background of muons produced in cosmic rays showers, the
telescopes look at the other side of the Earth, using it as a shield.
The increase in the range of muons at high energies (form kilometres
to several kilometres) together with the increase with energy of the
neutrino cross section give rise to an approximately exponential
increase of the effective areas of these devices in the GeV to PeV
energy range. Above a few TeV, the telescopes can determine the
direction of the incoming neutrinos with angular resolutions better
than 1$^\circ$, hence the name ``telescope''.  At energies above the
PeV, the Earth becomes opaque to neutrinos, but the atmospheric muon
flux decreases dramatically so that the neutrino telescopes can look
for downgoing neutrinos in that energy regime.  Other neutrino
flavours can be observed through the detection of hadronic or
electromagnetic showers or, in the case of tau neutrinos, via the
observation of its interaction and the subsequent decay of the
produced tau lepton.

\section{IceCube}
IceCube is at present the largest high energy neutrino telescope. It
is under construction at the South Pole and it will consist in its
final configuration of 80 + 6 strings~\cite{Karle09}. The 80 regular
strings contain 60 optical modules uniformly distributed between 1450
and 2450~m under the ice surface. The 6 remaining strings form the
so-called Deep Core on which the optical modules are located between
1760 and 2450~m, thus providing a denser distribution to yield a lower
energy threshold.  The optical modules contain 25~cm PMTs read by
waveform data acquisition electronics. In addition, the observatory
also contains a surface detector, IceTop, consisting in 80 pairs of tanks
on top of the strings, filled with ice and equipped with optical
modules. The completion of the telescope is expected to take place by
the beginning of 2011.  At present (2010) the dectector contains 79 of
the 86 strings and 73 of the 80 IceTop tank pairs.

A variety of studies have been carried out by the IceCube
collaboration during the construction phase. In this article we
briefly summarize some of the most recent results.

\begin{figure}[ht]
\vspace{12pt}
\hspace{0.2cm}
\includegraphics[scale=0.60]{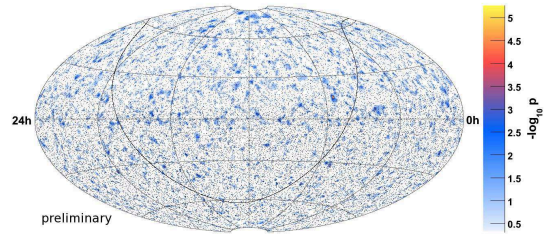}
\caption{Skymap obtained with 40-string IceCube. The black dots are
  the directionns of the selected events, the colour grade represents
  the pre-trial p-values.}
\label{IC40sky}
\end{figure}

The data taken by IceCube in the 40 string configuration in 2008 and
2009 corresponds to a live time of 375.5 days.  The final sample
contains almost 40 thousand events. Approximately one third of the
events are upgoing and the remaining two thirds downgoing. The
upgoing events, originating in the northern hemisphere, are mainly
atmospheric neutrinos in the ten to a few hundred TeV energy range,
while the downgoing events are high energy atmospheric muons above a
few PeV, where the atmospheric muon energy spectrum is softer than
that expected for a cosmic signal. Therefore, both sky hemispheres can
be analysed in the search for cosmic neutrinos. To this end, an
unbinned likelihood ratio based on the direction and estimated energy
of the tracks is used. The skymap of the selected events together with
the p-values obtained for each direction can be seen in
figure~\ref{IC40sky}. The highest significance (a p-value of 5.2
$\times$ 10$^{-6}$) is observed at a declination of 113.75$^\circ$ and
a right ascension of 15.15$^\circ$. Using scrambled data sets it is
estimated that such a significance or higher anywhere in the sky is
expected to take place 18\% of the time. This post-trial p-value
indicates that this fluctuation is compatible with the background.
  
\begin{figure}[htb]
\includegraphics[scale=0.30]{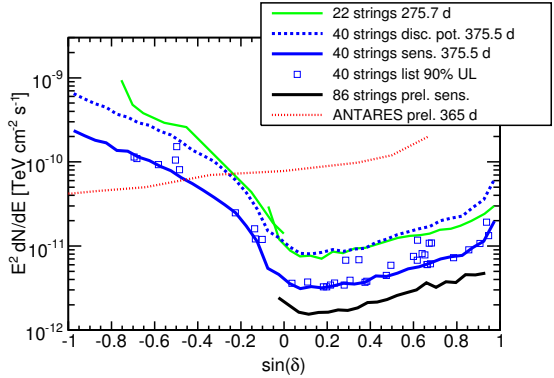}
\caption{Upper limits (90\% C.L.) and sensitivities.  The blue line
  shows the current IceCube 40 string upper limit and the squares the
  upper limits for a list of pre-selected
  sources~\cite{Aguilar10}. The previous 22-string IceCube 22 upper
  limit~\cite{Abbasi09} and the future 86-string configuration
  sensitivity are also shown.}
\label{ICpslimits}
\end{figure}

From the absence of a signal, upper limits on the neutrino flux from
point sources can be established assuming an E$^{-2}$ $\nu_{\mu}$
spectrum from the source.  The open blue squares in figure~\ref{ICpslimits}
show the 90\% C.L. upper limit for a candidate list of selected
sources~\cite{Aguilar10}. The sensitivities, i.e. the expected average
upper limits for a given direction, for IceCube 86 strings and ANTARES
(cf. section \ref{sec:antares}) are also shown. The current IceCube
sensitivity is three times better than the previous 22-string
sensitivity~\cite{Abbasi09}.
 
\begin{figure}[ht]
\includegraphics[scale=0.17]{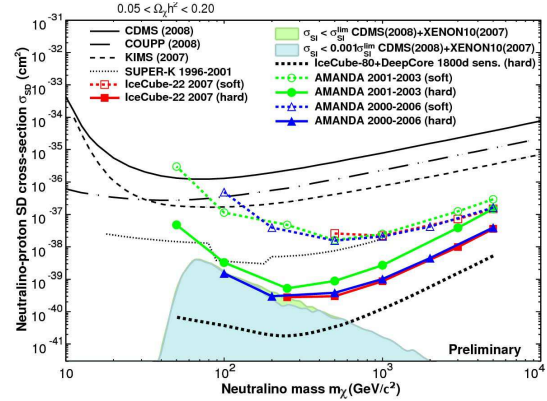}
\caption{
Neutralino-proton spin dependent cross-section as a function of neutralino
mass. The solid and dashed blue lines give the upper limits set by IceCube
22-string for soft (via $b \bar{b}$) and hard ($W^+ W^-$) annihilation 
channels. Also given are the limits set by direct search experiments. The
prediction for IceCube plus Deep Core after 5 years of data taking is also
shown (bottom dotted line).}
\label{ICdarkmatter}
\end{figure}

Neutrino telescopes can indirectly search for dark matter particles
looking for high energy neutrinos coming from the Sun. Weakly
interacting massive particles (WIMPs) can accumulate gravitationally
at the centre of astrophysical objects like the Sun, the Earth or the
Galactic Centre, and annihilate producing neutrinos. IceCube is
competitive with direct search experiments for WIMPs with mostly
spin-dependent interactions. As can be seen in
figure~\ref{ICdarkmatter}, the upper limit set by IceCube
20-string~\cite{AbbasiDM} is the most stringent for masses above
250~GeV provided the annihilation proceeds through the hard channels
($W^+ W^-$). Also shown are the limits set by AMANDA, IceCube's
forerunner, and the prediction for the sensitivity reached by the full
IceCube, including Deep Core, after five years of observation of the
Sun. As can be seen, it is more than one order of magnitude better and
covers a mass region from 50~GeV to several tens of TeV.
  
\section{Neutrino telescopes in lake and sea water}
The first attempt to use sea water as the medium in a neutrino
telescope was DUMAND\cite{Babson90,Roberts92}. Althought the project
was discontinued in 1996 and only a line was deployed (out of the nine
which DUMAND-II was planned to have) this initiative was the seed
for subsequent cosmic neutrino telescopes in water.

The BAIKAL neutrino telescope NT200+ is deployed in Lake Baikal,
Russia, 3.6~km from shore at a depth of 1.1~km. NT200 consists of
eight strings, 72~m long, each with 24 pair of optical modules and is
running since 1998. The extension NT200+ contains three additional
outer strings, 140~m long with 12 optical modules each and is working
since 2005. In addition, in 2008 a km3-prototype string with new
technology was installed. The most stringent upper limit provided by
NT200 for a diffuse astrophysical flux of $\nu_e + \nu_{\mu} +
\nu_{\tau}$ assuming an E$^{-2}$ spectrum is E$^2 \, \Phi$=2.9
$\times$10$^{-7}$ GeV cm$^{-2}$ s$^{-1}$ in the 20~TeV to 20~PeV
energy range~\cite{baikal}.

The NESTOR project started in 1989. Its site is near Pylos in the
Greek Ionian coast at a depth of around 4000~m. The designed tower
consists of 12 titanium stars with six arms of 16~m radius with a
pair of optical modules at the end of each arm adding up to a total of
144 photomultipliers.  A first prototype consisting of one floor with
a 5~m radius star and holding twelve optical modules was deployed in
2003. Using more than 5 million triggers the zenith angle distribution 
of atmospheric muons and their flux was measured~\cite{nestor}.  
The NESTOR initiative is now part of the KM3NeT consortium 
(see section~\ref{sec:km3net}).

The NEMO project started in 1998. The selected site is 80~km from the
coast of Capo Passero in Sicily at a depth of 3500~m, but a shallower
site (2031~m) closer to the shore (25 km from Catania) is also used
for tests.  The NEMO concept for the detector unit is a tower of 16
storeys separated vertically by 40~m with a total tower length of
750~m. The storeys consist of 20~m arms with two OMs at each end. As
part of NEMO Phase-1 a prototype tower was deployed at the end of
2006. A small sample of downgoing muons was analyzed and its zenith
angle distribution measured~\cite{nemo}. The tower suffered some
technical problems and stopped working at the beginning of 2007. The
NEMO collaboration is feeding the wider KM3NeT consortium with new
ideas and technical solutions (see section~\ref{sec:km3net}).

 The most advanced initiative in the Mediterranean Sea, ANTARES, 
is dealth with in the following section.
 
\section{ANTARES}
\label{sec:antares}

ANTARES is located in the Mediterranean sea, 40~km off the French
coast near Toulon. 

\begin{figure}[htb]
\includegraphics[scale=0.35]{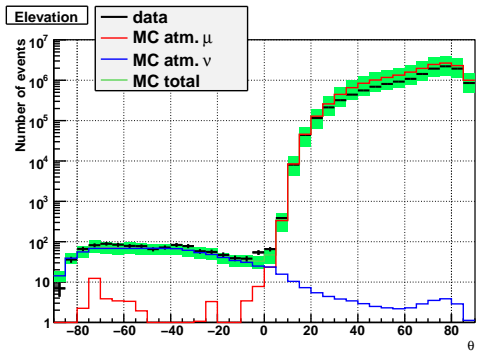}
\caption{
Elevation angle distribution for ANTARES events. Black points are the data,
the red and blue lines are the Monte Carlo predictions for atmospheric
muons and neutrinos. The green band indicates the total predicted 
atmospheric background together with its systematic uncertainty.}
\label{antelevation}
\end{figure}

ANTARES consists of 12 lines anchored to the sea bed. Each line contains 25
storeys.  The lowest storey is 100~m above the sea bed and the
vertical distance between consecutive storeys is 14.5~m.  The total
line length is 480~m.  Each storey has a triplet of OMs and an
electronics module. The OMs contain a 10-inch photomultiplier looking
45$^\circ$ downwards. The horizontal separation between lines is
60-80~m. The first line was deployed in 2006, the detector was
operated with 5-lines during several months in 2007 and was fully
deployed in 2008. Taking advantage of the possibility of detector
maintenance offered by water, the ANTARES collaboration has recovered
and repaired some of the lines and fixed problems in some of the
detector's interlink cables.

In figure~\ref{antelevation}, the distribution of the angle of
elevation above the horizon is shown for the data taken in 2007
(5-line configuration) and 2008 (9, 10 and 12 lines) corresponding to
a live time of 341 days. The total number of upgoing reconstructed
events is 1062. The Monte Carlo predictions for atmospheric muons and
neutrinos are also shown.

\begin{figure}[hbt]
\includegraphics[scale=0.70]{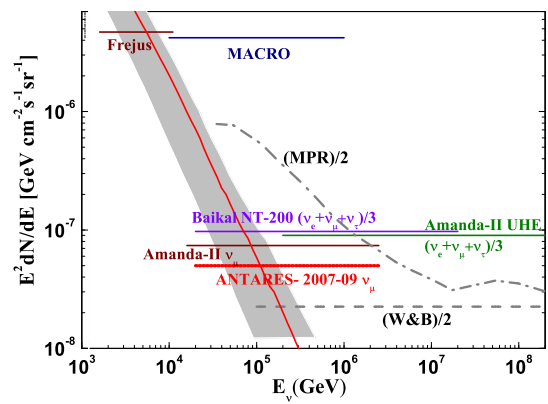}
\caption{ANTARES 90\% C.L. upper limit for an E$^{-2}$ high 
energy $\nu_{\mu }+\bar{\nu}_{\mu }$ diffuse flux. Upper limits
obtained by other experiments are also shown~\cite{otherdiffuse}.}
\label{antdiffuse}
\end{figure}

Using the data collected by ANTARES during the period from december
2007 to december 2009, a total live time of 334 days with different
detector configurations (9, 10 and 12 lines) a search for a diffuse
flux of astrophysical muon neutrinos was
performed~\cite{antdiffuse10}.  From the compatibility of the observed
number of events with the expected background and assuming an E$^{-2}$
flux spectrum for the signal, a 90\% C.L. upper limit on the $\nu_{\mu
}+\bar{\nu}_{\mu }$ diffuse flux of E$^2 \, \Phi$=5.3
$\times$10$^{-8}$ GeV cm$^{-2}$ s$^{-1}$ in the energy range 20~TeV to
2.5~PeV is obtained, as shown in figure~\ref{antdiffuse}

\begin{figure}[htb]
\includegraphics[scale=0.38]{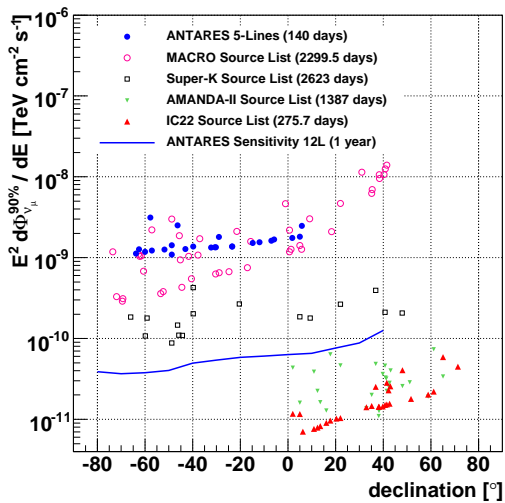}
\caption{Muon neutrino flux upper limit at 90\% C.L. obtained by the
  ANTARES 5-line telescope for a list 25 selected candidate sources
  (blue points) and expected sensitivity for the full ANTARES 12-line
  after one year of data taking. Also shown are the upper limits
  obtained by other experiments~\cite{otherpoint}.}
\label{antps5l}
\end{figure}

A first search for neutrino point sources has beeen performed using the 
5-line configuration data, corresponding to a live time of 140 days.
Binned and unbinned methods were used in the search~\cite{EM}. 
Figure~\ref{antps5l} shows the 90\% C.L. upper limits
obtained for a list of 25 candidate sources assuming an E$^{-2}$ spectrum.
Also shown is the sensitivity for ANTARES 12-line after one year of data
taking.

\section{KM3NeT}
\label{sec:km3net}
The ANTARES, NESTOR and NEMO collaborations have joined efforts in the
KM3NeT consortium to build a km$^3$-size neutrino telescope in the
Mediterranean Sea~\cite{km3net}.  The design and preparatory phases of
the project have been funded by the European Commission. At the end of
the design phase, a Conceptual Design Report that describes the goals
of the telescope and the options for its technical implementation was
delivered and a preliminary version of its Technical Design Report was
released in 2010~\cite{km3nettdr}. The minimum requirements for the
telescope are an effective volume of several km$^3$, sensitive to all
neutrino flavours, with an angular resolution better than 0.1$^\circ$
for muon neutrinos above 1~TeV and a target energy range 1-100~TeV for
point sources. 

\begin{figure}[htb]
\begin{center}
\begin{tabular}{cc}
\hspace{0 cm}
\includegraphics[scale=0.25]{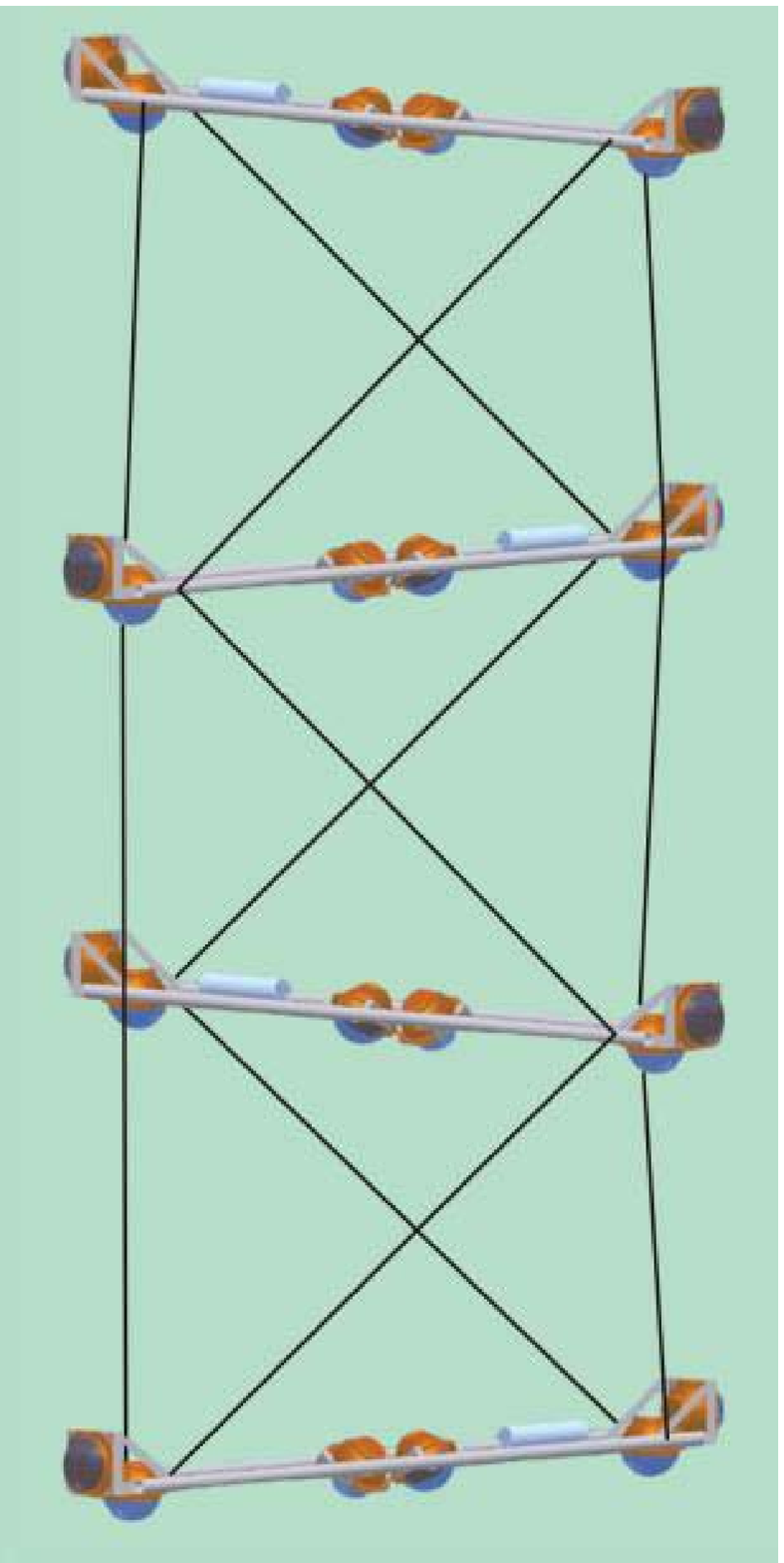}
&
\includegraphics[scale=0.17]{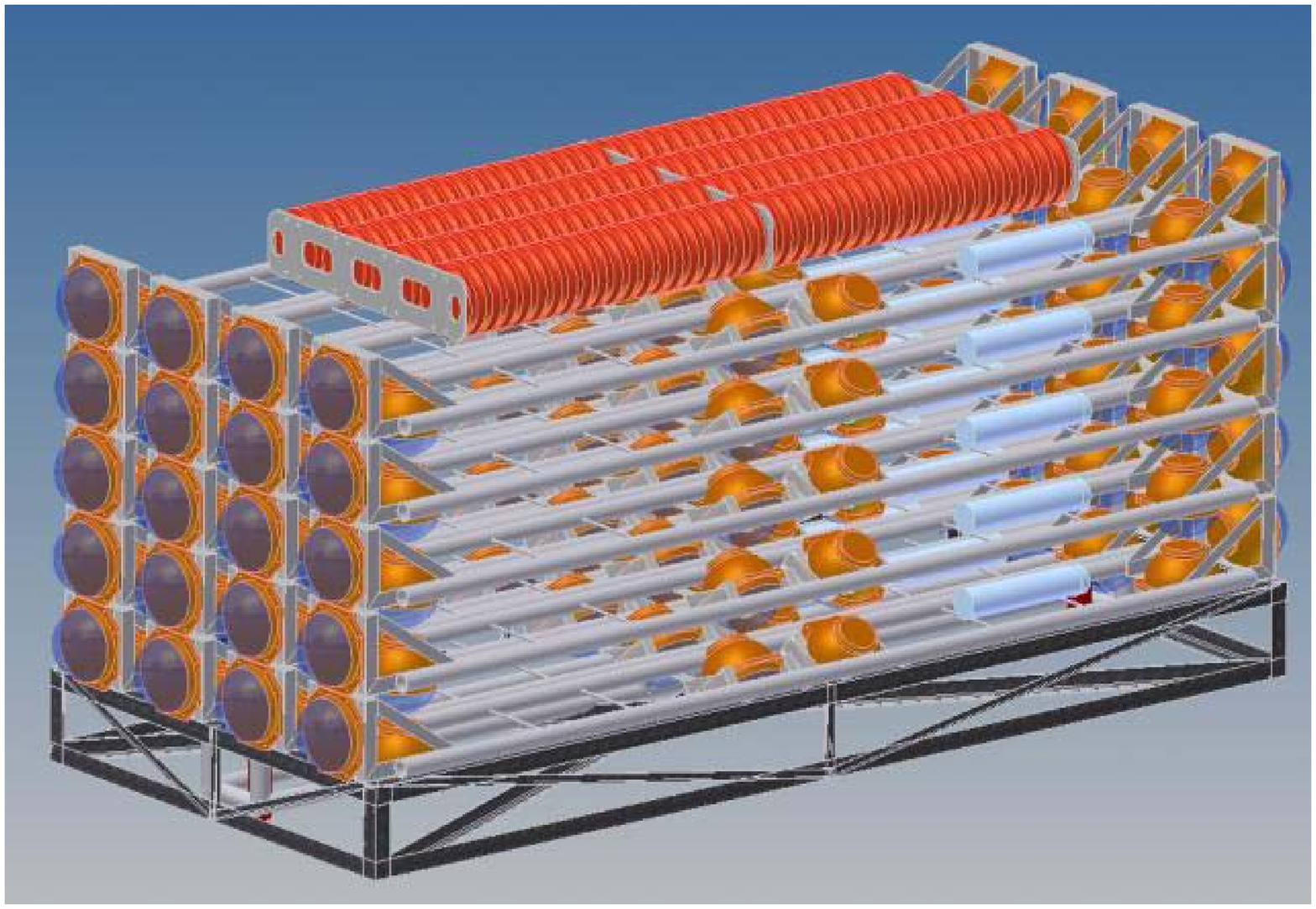}\\
\end{tabular}
\vspace{-0.1 cm}
\caption{A self-deployable tower proposed for KM3NeT. 
The optical modules in each storey are hold by bars. Each bar is perpendicular
to those in the nearby storeys. The tower can be densely packed for storage and
transportation and unfurls itself when it reaches the seabed.}
\label{KM3tower2}
\end{center}
\end{figure}
  
A wide variety of studies have been carried out to explore the
technical options to build and deploy such an instrument with an
overall cost below 250 Meuros. Among the different solutions proposed
let us mention as an example the concept of unfurlable tower (see
figure~\ref{KM3tower2}). This detector unit can be folded in a densely
packed structure suitable for storage, transportation and deployment
and it unfurls itself once it reaches the seabed. This enables the
deployment of several of these units during the same sea operation.  A
first test of the concept was succesfully carried out in 2010 with a
tower of reduced length.

KM3NeT is also meant to be a multidisciplinary infrastructure that
will house instrumentation relevant for seismology, radioactivity,
geomagnetism, oceanography, geochemistry and other sea sciencies,
involving therefore a large scientific community. KM3NeT is included
in the roadmap of the European Strategic Forum on Research
Infrastructures, is mentioned in the ASTRONET roadmap and was selected
by the ASPERA ERA-net as one of the seven astroparticle physics
infrastructures to be promoted~\cite{esfri}.

\section{Acknowledgements}
I would like to thank the organizers to invite me to NOW2010 in
Otranto.  I greatly enjoyed the nice scientific atmosphere and the
wonderful environment. This work has been partially supported by the
following funding agencies and grants: the Spanish Ministerio de
Ciencia y Tecnolog\'{\i}a, MICINN, FPA2099-13983-C02-01, ACI2009-1020
(KM3NeT) and Consolider MultiDark CSD2009-00064; Generalitat
Valenciana, Prometeo/2009/026; Commission of the European Communities,
KM3NeT Preparatory Phase, grant agreement 212525.

\end{document}